\documentstyle[12pt,epsf,epsfig,wrapfig]{article}
\begin{document}
\begin{center}
{\bfseries
ELECTROPRODUCTION OF A LIGHT NEUTRAL VECTOR MESON AT
NEXT-TO-LEADING ORDER}
\vskip 5mm
\underline{D.Yu. Ivanov}$^{1 \dag}$ and
L. Szymanowski$^{2}$
\vskip 5mm
{\small
(1) {\it
Institute of Mathematics, 630090 Novosibirsk, Russia
}
\\
(2) {\it
Soltan Institute for Nuclear Studies, Hoza 69, 00-681
Warsaw, Poland
}
\\
$\dag$ {\it
E-mail: d-ivanov@math.nsc.ru
}}
\end{center}
\vskip 5mm

\begin{abstract}
The process of a light neutral vector meson electroproduction is studied in
the framework of QCD factorization
in which the amplitude factorizes in a convolution of the nonperturbative
meson distribution amplitude  and the generalized parton densities
with the perturbatively calculable hard-scattering amplitudes.
We derive a complete set of hard-scattering amplitudes at next-to-leading
order (NLO) for the production
of vector mesons, $V=\rho^0, \omega, \phi$.
\end{abstract}

\vskip 8mm

\noindent
{\bf 1.}
The process of
elastic
neutral vector meson electroproduction on a nucleon,
\begin{equation}
\gamma^*(q)\, N(p) \to V(q^\prime)\, N(p^\prime)
\,\, , \quad \mbox{where} \quad
V=\rho^0,\,  \omega, \, \phi \, ,
\label{process}
\end{equation}
was studied in fix target
and in HERA collider experiments.
The primary motivation for the strong interest in this process
(and in the similar process of heavy quarkonium production) is that
it can potentially serve to constrain the gluon density in a nucleon
\cite{Ryskin:1992ui,BFGMS94}.
On the theoretical side, the large negative virtuality of the photon,
$q^2=-Q^2$, provides a hard
scale for the process which justifies the application of QCD factorization
methods that allow to separate the contributions to the amplitude
coming from different scales.
The factorization theorem \cite{CFS96} states that in a scaling limit,
$Q^2\to \infty$ and $x_{Bj}=Q^2/2(p\cdot q)$ fixed, a vector meson is produced
in the longitudinally polarized state by the longitudinally polarized photon
and that the amplitude of the
process (\ref{process}) is given by a convolution of the
nonperturbative
meson distribution amplitude (DA) and the generalized parton densities
(GPDs)
with the perturbatively calculable hard-scattering amplitudes.
In this contribution we present the results of our calculation of the
hard-scattering amplitudes at NLO.

\noindent{\bf 2.}
$p^2=p^{\prime 2}=m_N^2$ and $q^{\prime \, 2}=m^2_M$,
where $m_N$ and $m_M$ are a proton mass and a meson mass
respectively.
The invariant c.m. energy $s=(q+p)^2=W^2$. We define
\begin{eqnarray}
&&
\Delta=p^\prime -p \, , \ \ P=\frac{p+p^\prime}{2} \, , \ \ t=\Delta^2 \, ,
\nonumber \\
&&
(q-\Delta )^2=m^2_M \, , \ \ x_{Bj} =\frac{Q^2}{W^2+Q^2} \, .
\label{not1}
\end{eqnarray}

We introduce two light-cone vectors
\begin{equation}
n_{+}^2=n_{-}^2=0 \, , \ \ n_+ n_- = 1 \, .
\label{not2}
\end{equation}
Any vector $a$ is decomposed as
\begin{equation}
a^\mu=a^+n_+^\mu+a^-n_-^\mu+a_\perp \, , \ \ a^2=2\, a^+a^- - \vec a^2 \, .
\label{not3}
\end{equation}
We choose the light-cone
vectors in such a way that
\begin{eqnarray}
&&
p=(1+\xi)W n_+ + \frac{m_N^2}{2(1+\xi)W}\, n_- \, ,
\nonumber \\
&&
p^\prime=(1-\xi)W n_+
+\frac{(m_N^2+\vec \Delta^2)}{2(1-\xi)W}\, n_- +\Delta_\perp \, .
\label{not33}
\end{eqnarray}
We are interested in the kinematic region where
the invariant transferred momentum,
\begin{equation}
t=-\left(\frac{4\, \xi^2}{1-\xi^2}m_N^2+\frac{1+\xi
}{1-\xi}\vec \Delta^2\right) \, ,
\label{not4}
\end{equation}
is small, much smaller than $Q^2$.
In the scaling limit variable $\xi$ which parametrizes the
plus component of the momentum transfer equals $\xi=x_{Bj}/(2-x_{Bj} )$.

GPDs are defined as the matrix element of the light-cone quark
and gluon operators:
\begin{eqnarray}
&&
F^q (x,\xi,t)=\frac{1}{2}\int\frac{d\lambda}{2\pi}
e^{i x (P z)}\langle
p^\prime |\bar q \left(-\frac{z}{2}\right)\not \! n_-
q \left(\frac{z}{2}\right)
|p\rangle|_{z=\lambda n_-}
\nonumber \\
&&
=\frac{1}{2(Pn_- )}\left[
{\cal H}^q (x,\xi,t)\, \bar u(p^\prime)\not \! n_- u(p)+
{\cal E}^q (x,\xi,t)\, \bar
u(p^\prime)\frac{i\sigma^{\alpha\beta}n_{-\alpha}\Delta_\beta}{2\,m_N}  u(p)
\right] \, ,
\label{qGPD}
\end{eqnarray}
\begin{eqnarray}
&&
F^g (x,\xi,t)=\frac{1}{(Pn_-)}\int\frac{d\lambda}{2\pi}
e^{i x (P z)}\,
n_{-\alpha}n_{-\beta}\,\langle
p^\prime |G^{\alpha\mu} \left(-\frac{z}{2}\right)
G^{\beta}_\mu \left(\frac{z}{2}\right)
|p\rangle|_{z=\lambda n_-}
\nonumber \\
&&
=\frac{1}{2(Pn_- )}\left[
{\cal H}^g (x,\xi,t)\, \bar u(p^\prime)\not \! n_- u(p)+
{\cal E}^g (x,\xi,t)\, \bar
u(p^\prime)\frac{i\sigma^{\alpha\beta}n_{-\alpha}\Delta_\beta}{2\,m_N}
u(p)
\right] \, .
\label{gGPD}
\end{eqnarray}
In both cases the insertion of the path-ordered gauge factor between the
field operators is implied.
Momentum fraction $x$, $-1\leq x\leq 1$,  parametrizes parton momenta
with respect to the symmetric momentum $P=(p+p^\prime)/2$.
In the forward limit, $p^\prime=p$, the contributions proportional to
the functions ${\cal E}^q (x,\xi,t)$ and ${\cal E}^g (x,\xi,t)$ vanish,
the distributions ${\cal H}^q (x,\xi,t)$ and ${\cal H}^g (x,\xi,t)$ reduce
to the ordinary quark and gluon densities:
\begin{eqnarray}
&&
{\cal H}^q (x,0,0)=q(x) \ \ \mbox{for} \ \ x>0 \, ,
\nonumber \\
&&
{\cal H}^q (x,0,0)=-\bar q(-x) \ \ \mbox{for} \ \ x<0 \, ;
\nonumber \\
&&
{\cal H}^g (x,0,0)=\, x \, g(x) \ \ \mbox{for} \ \ x>0 \, .
\label{reduct}
\end{eqnarray}
Note that gluon GPD is even function of $x$,
${\cal H}^g (x,\xi,t)={\cal H}^g(-x,\xi,t)$.

The meson DA $\phi_V (z)$ is defined by the following relation
\begin{equation}
\label{Vdampl}
\langle0|\bar q(y) \gamma_\mu q(-y)|V_L(p)\rangle_{y^2\to 0}
 =p_{\mu}\, f_V
\int\limits^1_0 \!dz\, e^{i(2z-1)(py)} \phi_V (z)\, .
\end{equation}
It is normalized to unity $\int\limits^1_0 \phi_V (z) dz =1 \, $.
Here $z$ is a light-cone fraction of
a quark, $f_V$ is a meson dimensional coupling constant known from $V\to
e^+e^-$ decay, $f_\rho=198\pm 7 \, \rm{MeV}$.

\begin{eqnarray}
&&
 {\cal M}_{\gamma^*_L N\to V_L N}
=\frac{2\pi \sqrt{4\pi\alpha}\, f_V}
{N_c \, Q \, \xi} \int\limits^1_{0} dz \, \phi_V(z)
 \int\limits^1_{-1} dx
\Biggl[  \label{fact}\\
&&
\left.
 Q_V\left( T_g( z, x )\, F^g(x,\xi,t)+
T_{(+)} ( z, x)  F^{(+)} (x,\xi,t)
\right) +\sum_q e_q^V T_{q} ( z, x)  F^{q(+)} (x,\xi,t)
\right] \, .
\nonumber
\end{eqnarray}
Here the dependence of the GPDs, DA and the hard-scattering amplitudes
on factorization scale $\mu_F$ is suppressed for shortness.
Since we consider the leading
helicity non-flip amplitude, in eq. (\ref{fact})
the hard-scattering amplitudes do not depend on $t$.
The account of this dependence would lead to the power suppressed,
$\sim t/Q$, contribution.
$\alpha$ is a fine structure constant, $N_c=3$ is the number of QCD colors.
$Q_V$ depends on the meson flavor content. If one assumes it is
${1\over \sqrt{2}}(|u\bar u\rangle - |d\bar d\rangle )$,
${1\over \sqrt{2}}(|u\bar u\rangle + |d\bar
d\rangle )$ and $|s\bar s\rangle $ for $\rho$, $\omega$ and $\phi$
respectively, than $Q_\rho={1\over \sqrt{2}}$,
$Q_\omega={1\over 3\sqrt{2}}$ and $Q_\phi={-1\over 3}$,
the sum in the last term of (\ref{fact}) runs over $q=u,d$ for
$V=\rho,\omega$ and $q=s$ for $V=\phi$ and
$$
e_u^\rho=e_u^\omega={2\over 3\sqrt{2}}\, ,
\quad
e_d^\rho=-e_d^\omega={1\over 3\sqrt{2}}\,
\quad e_s^\phi={-1\over 3}\, .
$$
$F^{q(+)} (x,\xi,t)=
F^{q}(x,\xi,t)-F^{q}(-x,\xi,t)$ denotes a singlet quark GPD,
$F^{(+)} (x,\xi,t)=\sum_{q=u,d,s}F^{q(+)} (x,\xi,t)$ stands for the sum of
all light flavors.

Due to odd C- parity of a vector meson $\phi_V(z)=\phi_V(1-z)$. Moreover,
since $V$ and $\gamma^*$ have the same C- parities, $\gamma^*\to V$
transition selects the C-even gluon and
singlet quark contributions, whereas the
C-odd quark combination $F^{q(-)} (x,\xi,t)=
F^{q}(x,\xi,t)+F^{q}(-x,\xi,t)$ decouples in (\ref{fact}).

\noindent
{\bf 3.} Below we present the results of our calculation of the
hard-scattering amplitudes in the
$\overline{\rm{MS}}$ scheme.

$T_q(z,x)$
may be
obtained by the following substitution from the known results for a
pion EM formfactor, see also \cite{Belitsky:2001nq},
\begin{equation}
\label{qqq}
T_q( z, x )=\left\{T\left(z,\frac{x+\xi}{2\xi} -i\epsilon \right)-
T\left(\bar z,\frac{\xi-x}{2\xi} -i\epsilon \right) \right\}+
\left\{z\to \bar z \right\} \, ,
\end{equation}
\begin{eqnarray}
\label{PiEM}
&&
T(v,u)= \frac{\alpha_S(\mu_R^2) C_F}{4vu}
\left( 1+\frac{\alpha_S(\mu_R^2)}{4\pi}\left[
c_1\left(2\left[3+\ln (vu)\right]
\ln{\left(\frac{Q^2}{\mu^2_F}\right)}+\ln^2 (vu)
\right.\right.\right.
\nonumber \\
&&
\left.
+
6\ln (vu)
-\frac{\ln (v)}{\bar v}-\frac{\ln (u)}{\bar u}-\frac{28}{3}\right) +
\beta_0\left(\frac{5}{3}-\ln (vu)-\ln{\left(\frac{Q^2}{\mu^2_R}\right)}
\right)
\nonumber \\
&&
+c_2\left(
2\frac{(\bar v v^2+\bar u u^2)}{(v-u)^3}
\left[
Li_2(\bar u)-Li_2(\bar v)+Li_2(v)-Li_2(u)+\ln(\bar v)\ln(u)-\ln(\bar
u)\ln(v)
\right] \right. \nonumber \\
&&
+2\left[
Li_2(\bar u)+Li_2(\bar v)-Li_2(u)-Li_2(v)+\ln(\bar v)\ln(u)+\ln(\bar
u)\ln(v)
\right]
\nonumber \\
&&
\left.\left.\left.
+4\frac{vu\ln (vu)}{(v-u)^2}+2\frac{(v+u-2vu)\ln{\bar v\bar u}}{(v-u)^2}
-4\ln(\bar v)\ln(\bar u)-\frac{20}{3}
\right)
\right]
\right) \ .
\end{eqnarray}
Here and below we use a shorthand notation $\bar u=1-u$ for any light-cone
fraction. $\mu_R$ is a renormalization scale for a strong coupling,
$\beta_0=\frac{11N_c}{3}-\frac{2n_f}{3}$, $n_f$ is the effective number of
quark flavors. $C_F=\frac{N_c^2-1}{2N_c}$,
$Li_2(z)=-\int\limits^z_0\frac{dt}{t}\ln(1-t)$. Also we denote
\begin{equation}
c_1=C_F \, , \quad c_2=C_F-\frac{C_A}{2}=-\frac{1}{2N_c}\, .
\label{colconst}
\end{equation}

$T_{(+)}( z, x )$ starts from NLO
\begin{equation}
\label{q+}
T_{(+)}( z, x )=\frac{\alpha_S^2(\mu_R^2)\, C_F}{(8\pi) z\bar z}\, {\cal
I}_q\left(z,\frac{x-\xi}{2\xi}+i\epsilon\right) \ ,
\end{equation}
here
\begin{eqnarray}
\label{quarktogluon}
&& {\cal I}_q(z,y)  =
\left\{
\frac{2y+1}{y(y+1)}
\left(
\left(\ln\biggl(\frac{Q^2}{\mu_F^2}\biggr) -1 +\ln(z)\right)
\Bigl(y\ln(-y)-(y+1)\ln(y+1)
\Bigr)\right.
\right.
\nonumber \\
&&
\left.  +\frac{y}{2}\ln^2(-y)-\frac{y+1}{2}
\ln^2(y+1)
\right) +
\frac{y\ln(-y)+(y+1)\ln(y+1)}{y(y+1)}
\nonumber \\
&&
\left. -\frac{R(z,y)}{y+z} +  \frac{y(y+1)+(y+z)^2}{(y+ z)^2}\, H(z,y)
\right\}+\left\{z\to \bar
z\right\} \, ,
\end{eqnarray}
where we introduced two auxiliary functions
\begin{equation}
\label{R}
R(z,y)=z \ln (-y) +
   \bar z\ln (y+1) + z\ln (z) +
  \bar z \ln (\bar z) \, ,
\end{equation}
\begin{equation}
\label{H}
H(z,y)=
 Li_2(y+1)-Li_2(-y)+Li_2(z)-Li_2(\bar z)
+ \ln (-y)\,\ln (\bar z) -
  \ln (y+1)\,\ln (z) \, .
\end{equation}

For the gluonic contribution we obtain
\begin{equation}
\label{gluon}
T_{g}( z, x )=\frac{\alpha_S(\mu_R^2) \, \xi}{z\bar z \,
(x+\xi-i\epsilon)(x-\xi+i\epsilon)}
\left[1+
\frac{\alpha_S(\mu_R^2)}{4\pi}\, {\cal
I}_g\left(z,\frac{x-\xi}{2\xi}+i\epsilon\right)\right] \ ,
\end{equation}
where
\begin{eqnarray}
\label{Ig}
&&
{\cal I}_g(z,y)=\left\{
-\frac{\beta_0}{2}\left(\ln\biggl(\frac{Q^2}{\mu_R^2}\biggr)-1\right)
\right.
\\
&&
+
\left(\ln\biggl(\frac{Q^2}{\mu_F^2}\biggr)-1\right)
\left[\frac{c_1}{2}
\left(\frac{y\ln(-y)}{y+1}+\frac{(y+1)\ln(y+1)}{y} \right)
+c_1\left(
\frac{3}{2}+2z\ln(\bar z)\right) \right.
\nonumber \\
&&
\left.
+\frac{2(c_1-c_2)
(y^2+(y+1)^2)}{ y(y+1)}\Bigl(y\ln(-y)-(y+1)\ln(y+1)\Bigr)
+\frac{\beta_0}{2} \right]
\nonumber \\
&&
+
(c_1-c_2)(2y+1)\Bigl(\ln(-y)-\ln(y+1)\Bigr)
\left(\frac{3}{2}+\ln(z\bar z)+\ln(-y)+\ln(y+1)\right)
\nonumber \\
&&
\left.
+\left(c_1(y(y+1)+(y+z)^2)-c_2(2y+1)(y+z)\right)\left[
\frac{y(y+1)+(y+z)^2}{(y+z)^3}H(z,y)
\right.\right.
\nonumber \\
&&
\left.\left.
-\frac{R(z,y)}{(y+z)^2}+\frac{\ln(-y)-\ln(y+1)+\ln(z)-\ln(\bar z)}{2(y+z)}
\right] -\frac{c_1(2y+1)R(z,y)}{2(y+z)} -2c_1   \right.
\nonumber \\
&&
\left.
-(c_1-c_2)\Bigl(\ln(z\bar z)-2\Bigr)
\left(
\frac{y\ln(-y)}{y+1}+\frac{(y+1)\ln(y+1)}{y}
\right)
+c_1(1+3z)\ln(\bar z)
\right.
\nonumber \\
&&
-\frac{(3c_1-4c_2)}{4}\left(
\frac{y\ln^2(-y)}{y+1}+\frac{(y+1)\ln^2(y+1)}{y}\right)
+ c_1 z\ln^2(\bar z)
\left. \right.
\nonumber \\
&&
+\Bigl(\ln(-y)+\ln(y+1)\Bigr)\left(c_1\left(\bar z\ln(z)-\frac{1}{4}\right)
+2c_2\right)
\Biggr\}+\left\{z\to \bar z\right\} \, .
\nonumber
\end{eqnarray}

\noindent
{\bf 4.} Above equations give a complete description of a neutral
vector meson electroproduction with NLO accuracy.
At leading order we reproduce
known result \cite{Mank97}, our results for the NLO correction are new.

At high energies, $W^2\gg Q^2$, the imaginary part of the amplitude
dominates. Leading contribution to the NLO correction comes from the
integration region $\xi\ll |x|\ll 1$, simplifying
the gluon hard-scattering amplitude in this limit we obtain the estimate
\begin{eqnarray}
&&
 {\cal M}_{\gamma^*_L N\to V_L N}
\approx \frac{-2\, i\, \pi^2 \sqrt{4\pi\alpha}\, \alpha_S f_V Q_V}
{N_c \, Q \, \xi}
 \label{appr}\\
&&
 \int\limits^1_{0}\frac{ dz \, \phi_V(z)}{z\bar z}\Biggl[
\left.
F^g(\xi,\xi,t)+\frac{\alpha_S N_c}{\pi}
\ln\left(\frac{Q^2z\bar z}{\mu_F^2}\right)
\int\limits^1_{\xi} \frac{dx}{x}  F^g(x,\xi,t)
\right] \, .
\nonumber
\end{eqnarray}
Given the behavior of the gluon GPD at small $x$, $F^g(x,\xi,t)\sim const$,
we see that NLO correction is parametrically
large and negative unless one chooses the
value of the factorization scale sufficiently lower than the kinematic
scale. For the asymptotic form of meson DA, $\phi^{as}_V(z)=6z\bar z$,
the last term in (\ref{appr}) changes the sign at $\mu_F=\frac{Q}{e}$,
for the DA with a more broad shape this happens at even lower values of
$\mu_F$.

\vspace*{1cm}

\centerline{
{\bf  Acknowledgments}}

\vspace*{0.5cm}

\noindent Work of D.I. is supported in part by Alexander von
Humboldt Foundation and by grants DFG 436, RFBR 03-02-17734, L.Sz.
is partially supported by the French-Polish scientific agreement
Polonium.


\begin{thebibliography}{99}

\bibitem{Ryskin:1992ui}
M.G.~Ryskin,
Z.\ Phys.\ C {\bf 57} (1993) 89.

\bibitem{BFGMS94}
S.J.~Brodsky at al.,
Phys.\ Rev.\ D {\bf 50} (1994) 3134.

\bibitem{CFS96}
J.C.~Collins, L.~Frankfurt and M.~Strikman,
Phys.\ Rev.\ D {\bf 56} (1997) 2982.

\bibitem{Belitsky:2001nq}
A.V.~Belitsky and D.~Muller,
Phys.\ Lett.\ B {\bf 513} (2001) 349.


\bibitem{Mank97}
L.~Mankiewicz, G.~Piller and T.~Weigl,
Eur.\ Phys.\ J.\ C {\bf 5} (1998) 119.

\end{thebibliography}
\end{document}